\documentclass[a4paper]{article}
\usepackage{epsfig,amsmath,amsfonts,amssymb,setspace,multirow,textcomp}
\usepackage[T1]{fontenc}
\makeatletter
\renewcommand{\@evenfoot}{\hfil \thepage \hfil}
\renewcommand{\@oddfoot}{\hfil \thepage \hfil}
\makeatother

\renewenvironment{thebibliography}[1]{\begin{oldthebibliography}{#1}\setlength{\parskip}{0ex}\setlength{\itemsep}{0ex}}{\end{oldthebibliography}}

\begin{document}
\fontsize{11}{11}\selectfont 
\title{Small-scale variability in the spectrum of Vega}
\author{\textsl{S.\,M.~Pokhvala, B.\,E.~Zhilyaev}} 
\date{\vspace*{-6ex}}
\maketitle
\begin{center} {\small $Main Astronomical Observatory, NAS \,\, of Ukraine, Zabalotnoho \,27, 03680, Kyiv, Ukraine$}\\ 
{\tt nightspirit10@gmail.com}
\end{center}

\begin{abstract}
We reported the results of observations of small-scale variability in the hydrogen Balmer lines in Vega. Spectral observations were carried out with low-resolution spectrograph (R $\simeq$ 600) installed in the Main Astronomical Observatory, Ukraine. Spectra were obtained with a time resolution in the second range. It has been found that Vega shows variations in the hydrogen lines $H_{\beta} $, $H_{\gamma} $, $H_{ \delta} $. This can be interpreted that their variations are non-radial pulsations. The characteristic time of the observed variations ranges from 300 to 1200 sec. The horizontal scale for oscillating elements is about 800 Mm, which is comparable to the solar radius. The radial velocity of the variations is about 36 km/s.


\end{abstract}

\section*{\sc introduction}

\indent \indent Vega is a low-amplitude variable star. Early claims to photometric variability in Vega were largely the work: of Guthnick \cite{Guthnick_1}, \cite{Guthnick_2}, Fath \cite{Fath}, and Fernie \cite{Fernie}. We summarize Fernie's summations below. 

It was concluded that Vega shows a variability of about 0.03 mag on a time scale of hours and varies by 0.08 mag on a scale of months. 
Percy \cite{Percy}, suspected variability on a scale of about 0.02 mag. 

The results of 14 nights of observation were obtained.  
The conclusion seems to be,  that for most of the four months, Vega was constant to within $\pm$ 0.006 mags, but that on at least one occasion it brightened
by 0.04 mag.

If Vega is variable, what is the expected time scale? As a basis for discussion, one shall consider the fundamental period for radial oscillation. This is determined almost entirely by the star's mass and radius, which can be derived reliably in a well-studied star like Vega. The radius of Vega was obtained to R = 2.80 $\pm$ 0.17 $R_{\odot}$.
Adopting $T_{e}$ = 9600 $\pm$ 200 K and $M_{bol}$ = + 0.2 $\pm$ 0.2, give masses of 2.1 $\pm$ 0.3 $M_{\odot}$ for Vega.

The period of the fundamental radial pulsation can be calculated from the relation P$\surd \rho $ = Q, where $\rho $ denotes the mean stellar density. Q   is a function of $M_{*}/R_{*}$. Cogan \cite{Cogan} created models whose range of masses and radii includes Vega. It is Q = 0.0335 $\pm$  0.0005 days. Given this value and the mass and radius, a P = 0.107 $\pm$ 0.013 days can be predicted (9245 sec).

Rotation period  is 0.71 $\pm$  0.02 days, Monnier \cite{Monnier}.

\section*{\sc Observations}

\indent \indent Spectral observations of Vega were carried in August 2024 in the Main Astronomical Observatory.  The goal of the observations was to obtain spectra of the star, in order to study fast variation in the spectral lines. To investigate the spectral variability of Vega spectra were obtained at 1 hour intervals using a spectrograph with spectral resolution R $\sim$ 200 and R $\sim$ 600. In the first case  a time resolution was  0.790 seconds, and spectral resolution $\triangle \lambda$ = 5.595 \AA/pixel. In the second case a time resolution was  0.880 seconds, and spectral resolution $\triangle \lambda$ = 3.297 \AA/pixel. Fig 1 shows recorded spectra of Vega.


\section*{\sc The detection of line-profile variations}

The following Fig. 2 shows the dependence of the variance from the mean intensity in the spectrum of the star. For Poisson random variables, the following relation holds \cite{Pollard79}:
$$ (n-1)\,\sigma^{2}/mean = \chi^{2}_{2}$$
where $\sigma$ is the variance, $n$ is the number of measurements. The formula allows setting a detection threshold by using the $\chi^{2}_{2}$ probability distribution. For $n$ = 854 the 98\% detection threshold for relative variations is 0.04. 

In Fig. 2 one can see activity in the hydrogen lines $H_{\alpha} $, $H_{\beta} $, $H_{\gamma} $, and Mg b 5167, 5173,5184 \AA \, lines, as well as atmospheric lines of oxygen ($\lambda$ 6870 \AA \, and $\lambda$ 7600 \AA) and water ($\lambda$ 7180 \AA).  

The work solved the problem of finding the intrinsic variability in the lines of the spectrum of stars, by taking into account the difference in the intrinsic variability spectra and the noise distribution spectrum. 

\begin{figure}[!h]
\centering
\begin{minipage}[t]{.45\linewidth}
\centering
\epsfig{file = 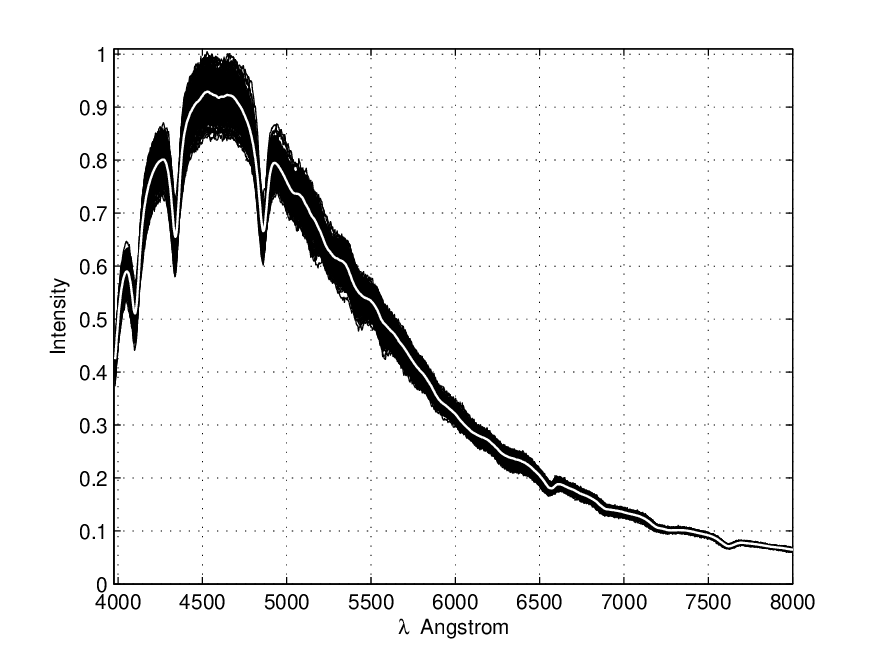,width = 1.0\linewidth} \caption{The spectra of Vega.}\label{fig1}
\end{minipage}
\hfill
\begin{minipage}[t]{.45\linewidth}
\centering
\epsfig{file = 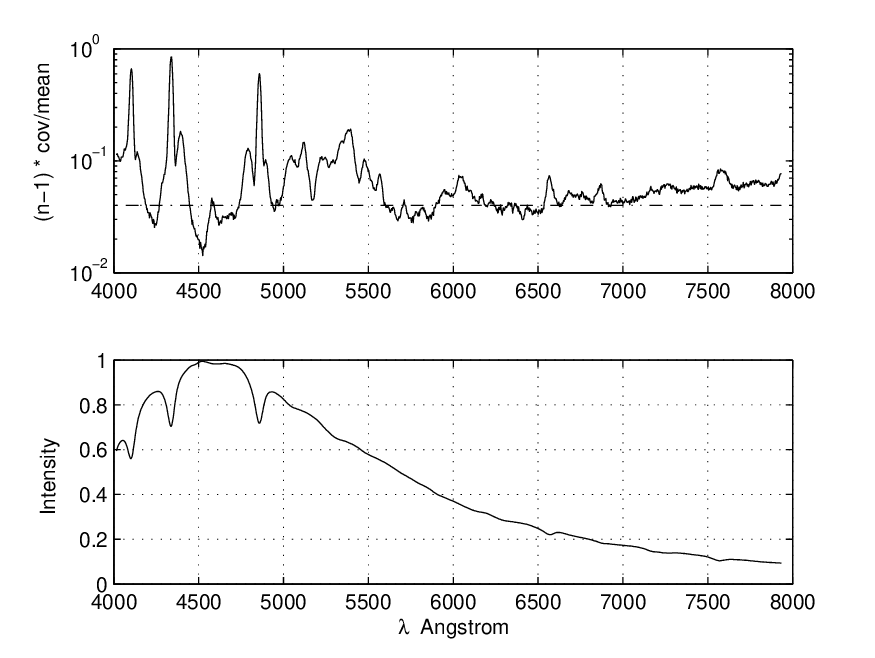,width = 1.0\linewidth} \caption{Variations in the spectrum of Vega.}\label{fig2}
\end{minipage}
\end{figure}


We use the Phase Discrimination to free the light curve of the spectral lines from spurious harmonics provoked by interference due to variations in atmospheric transparency and guiding errors. These intensity variations can be much larger in amplitude than the intrinsic variations in the spectral lines that are the subject of study. The Phase Discrimination Technique is described in detail in Zhilyaev \& Pokhvala \cite{Zhi2020}. 


\begin{figure}[!h]
\centering
\begin{minipage}[t]{.45\linewidth}
\centering
\epsfig{file = 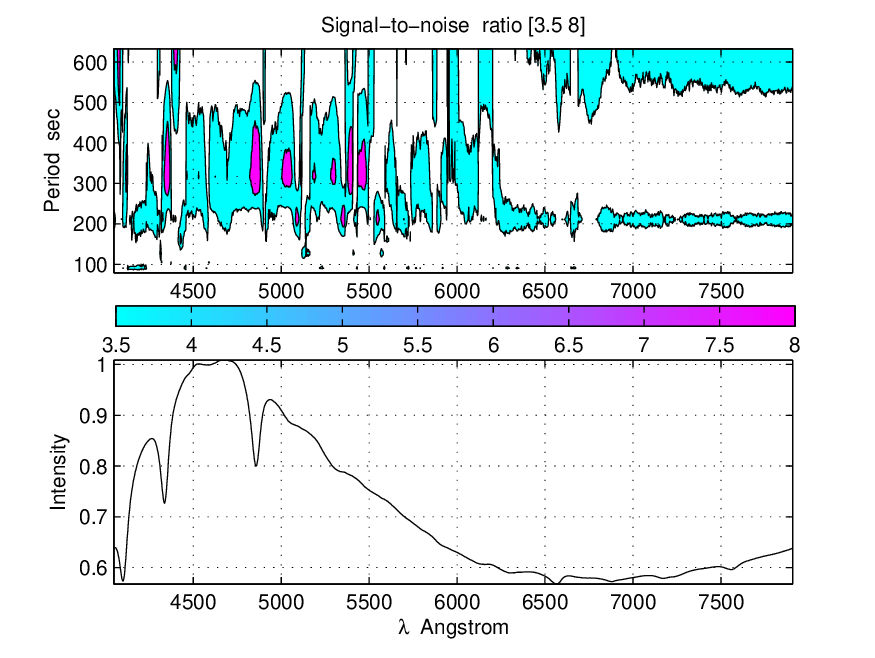,width = 1.0\linewidth} \caption{The map of spectra of Vega.}\label{fig3}
\end{minipage}
\hfill
\begin{minipage}[t]{.45\linewidth}
\centering
\epsfig{file = 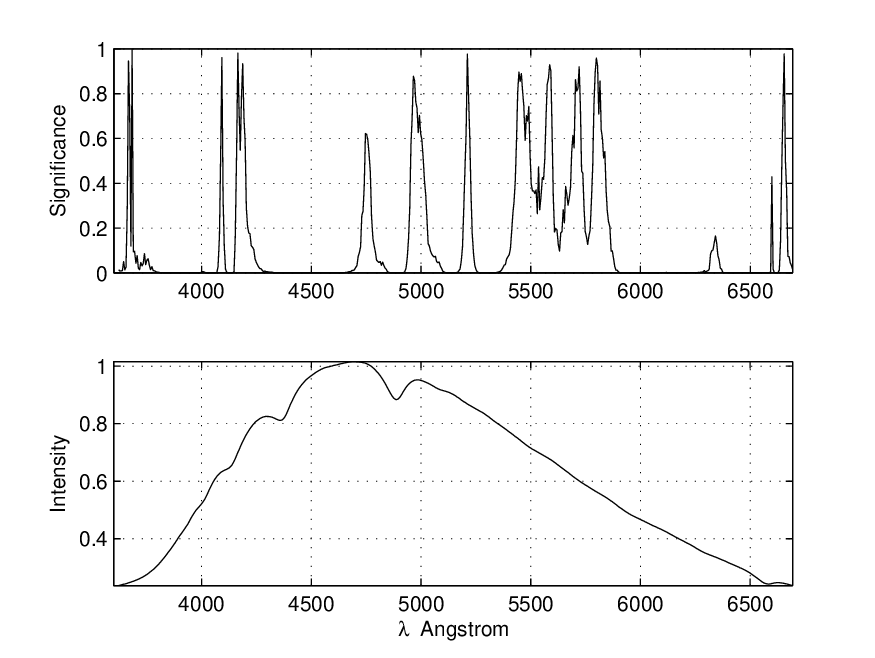,width = 1.0\linewidth} \caption{Significance of variations in the spectrum. The detection threshold of activity.}\label{fig4}
\end{minipage}
\hfill
\end{figure}


We calculate the phase spectra of the signal and the interference. If we assume that the phase error is $\pm$ 1 radian, this provides a criterion for harmonic selection. Harmonics falling within the phase difference corridor of $\pm$ 1 radian are treated as harmonics must be separated.

The interference signal is a "white light curve" which is obtained by integrating the spectrum over the operating wavelength range. After freeing the light curve of each spectral line from spurious harmonics, we select harmonics in the Fourier transform whose signal-to-noise ratio exceeds a critical level.  This allows us to recover the noise-free light curve of the spectral line.

In Fig. 3 one can see activity in the hydrogen lines $H_{\alpha} $, $H_{\beta} $, $H_{\gamma} $, $H_{ \delta} $, and Mg b 5167, 5173,5184 \AA \, lines. 

Atmospheric lines of oxygen ($\lambda$ 6870 \AA \, and $\lambda$ 7600 \AA) and water ($\lambda$ 7180 \AA) demonstrate a single peak variability structure (Fig. 3) with periods different from those of the hydrogen lines.

Activity detection occurs on a level that, on average, crosses one peak of noise. These lines show several activity islands with different periods. It is important to note that not all lines are active. This serves as a serious argument in favor of the proposed method for detecting and estimating the spectral variability of stars. 


Fig. 4 can be used to quantify the distribution of statistically significant variability as a function of position within a line profile. In principle, the morphology of this distribution provides insight into the nature of the variability \cite{Fullerton96}. 

Line profile variations are a very valuable diagnostic to detect both radial and non-radial oscillations (e.g. \cite{Aerts03} and references therein; \cite{Hekker06}), and to characterize the wave numbers $(l,m)$ of such self-excited oscillations. It is worthwhile to detect them for Vega with confirmed oscillations and to use them for empirical mode identification. 

The variations of the line profile are demonstrated by Figure 5. Figure 6 shows the variations of the equivalent $H_{\beta}$ line width and radial velocity. The Fourier spectra of these variations show a clear peak with a period of about 300 sec (Figs. 7 and 8).

The dynamic spectrum of variation of the $H_{\beta}$ line profile is shown in Fig. 9. Figure 10 shows the light curves of the $H_{\beta}$ line center and the nearby continuum. It is easy to see a harmonic with a period of 300 sec in the center of the line with an amplitude of 0.01 stellar magnitude.





\begin{figure}[!h]
\centering
\begin{minipage}[t]{.45\linewidth}
\centering
\epsfig{file = 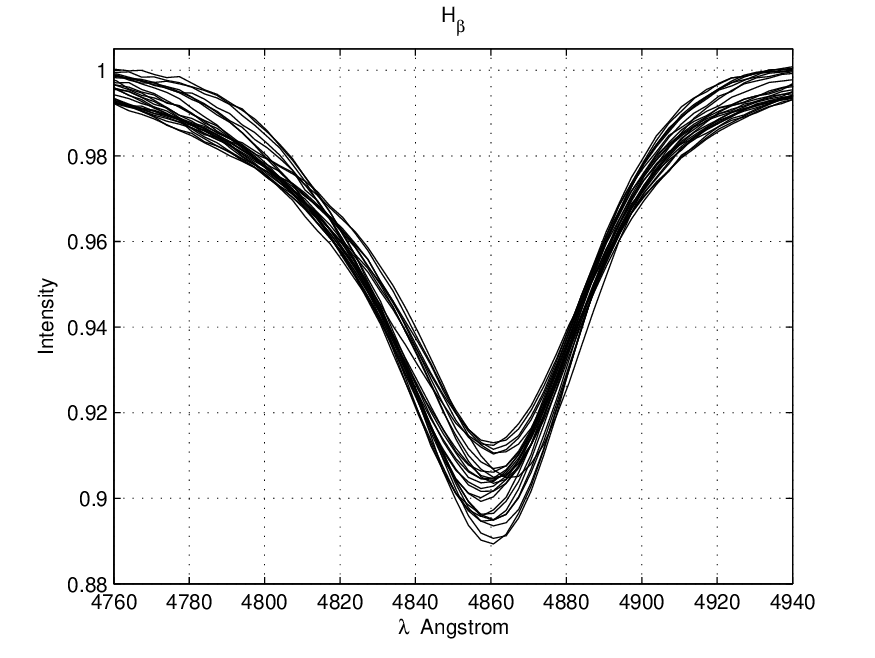,width = 1.0\linewidth} \caption{$H_{\beta}$ lines in Vega's spectrum.}\label{fig5}
\end{minipage}
\hfill
\begin{minipage}[t]{.45\linewidth}
\centering
\epsfig{file = 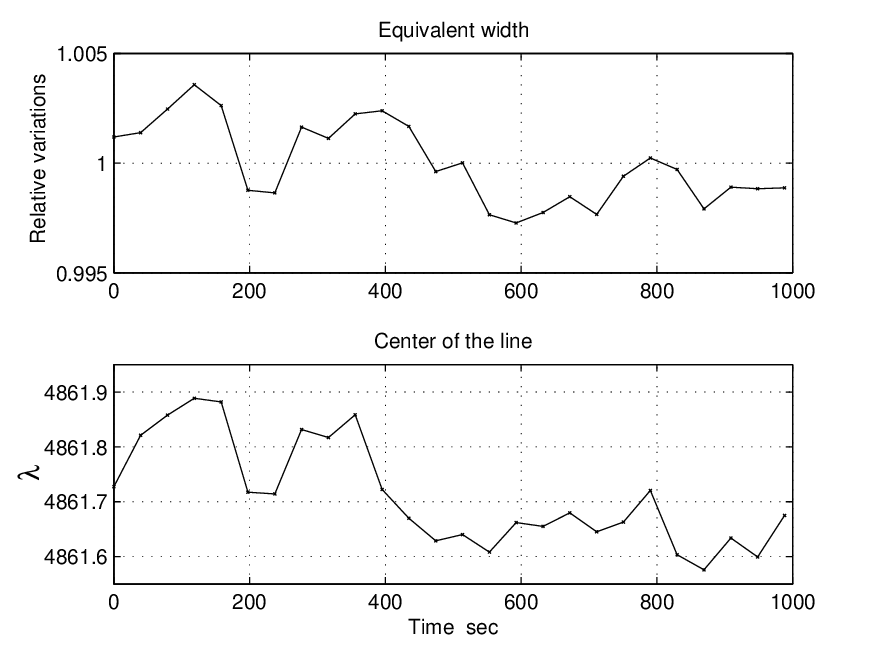,width = 1.0\linewidth} \caption{Variations in equivalent width (top panel) and center of the $H_{\beta}$ line (bottom panel).}\label{fig6}
\end{minipage}
\hfill
\end{figure}
%


\begin{figure}[!h]
\centering
\begin{minipage}[t]{.45\linewidth}
\centering
\epsfig{file = 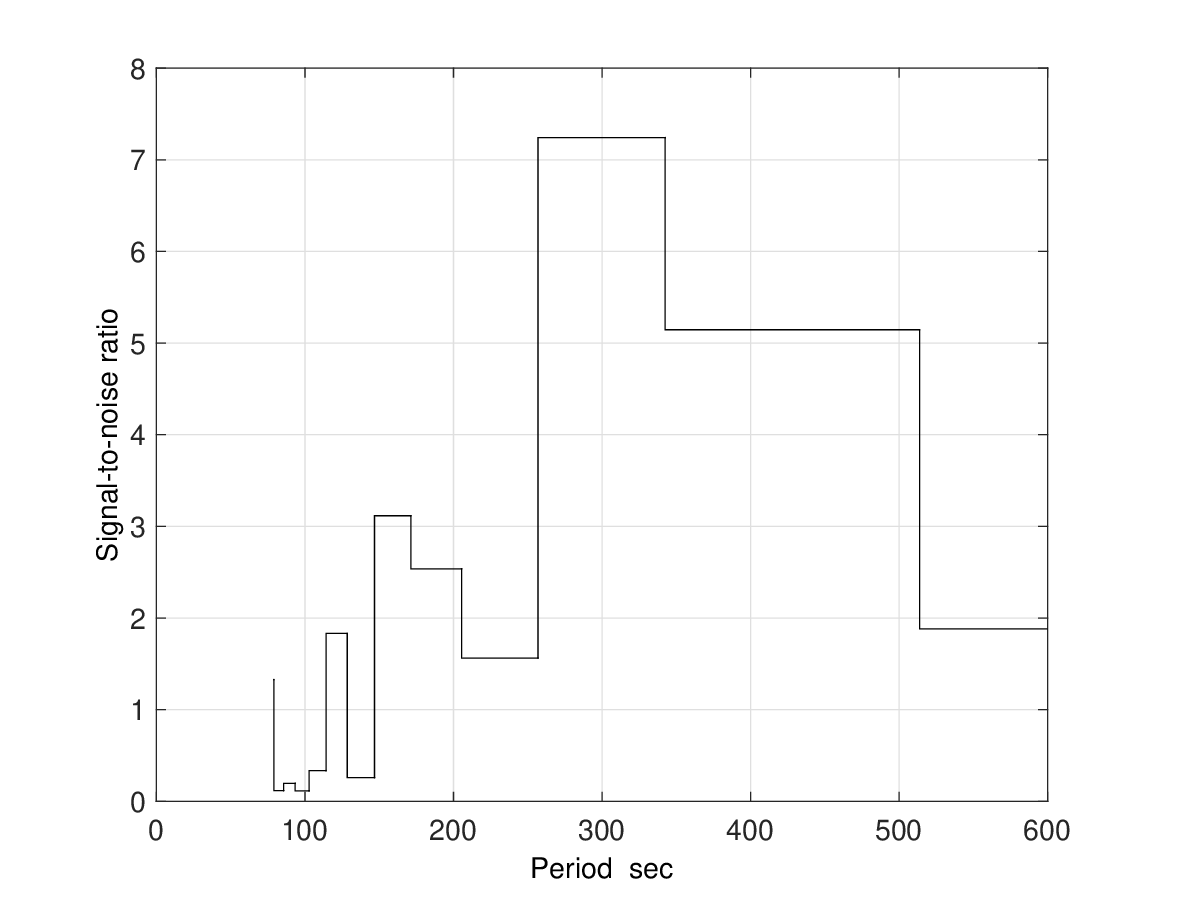,width = 1.0\linewidth} \caption{The center of the $H_{ \beta }$ line. Fourier spectrum of variations.}\label{fig7}
\end{minipage}
\hfill
\begin{minipage}[t]{.45\linewidth}
\centering
\epsfig{file = 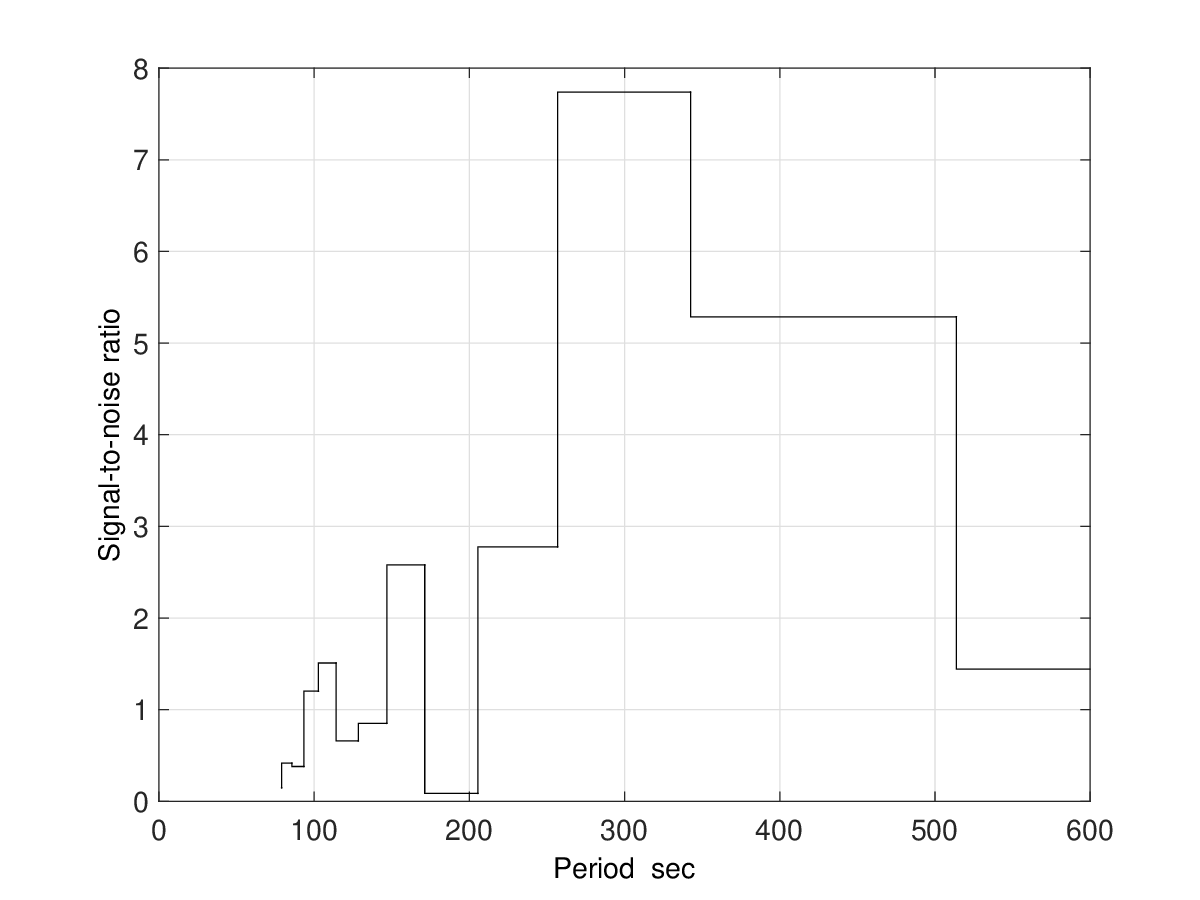,width = 1.0\linewidth} \caption{Fourier spectrum of the light curve of the line center.}\label{fig8}
\end{minipage}
\hfill
\end{figure}
%


\begin{figure}[!h]
\centering
\begin{minipage}[t]{.45\linewidth}
\centering
\epsfig{file = 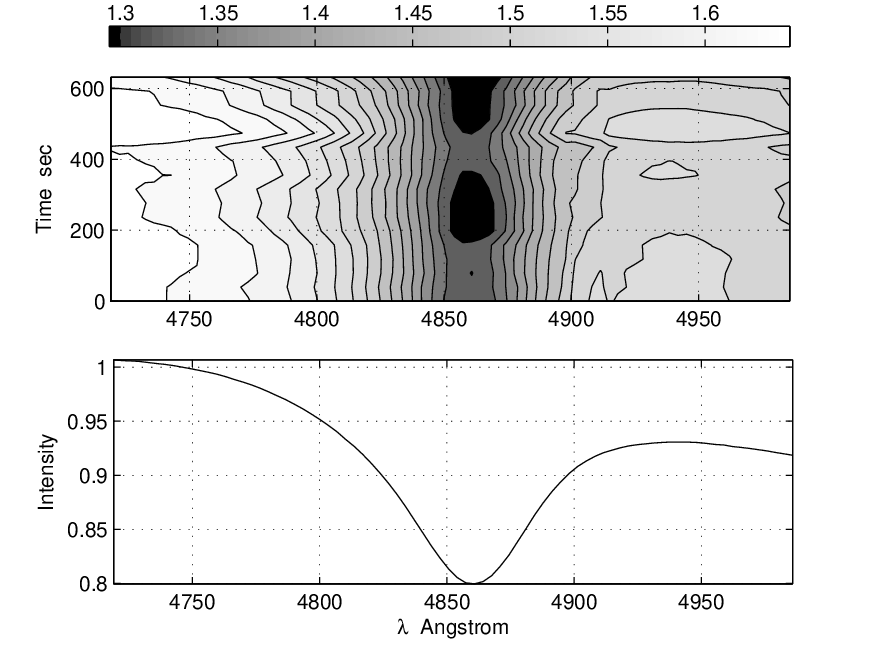,width = 1.0\linewidth} \caption{Dynamic spectra for the $H_{\beta}$ line.}\label{fig5}
\end{minipage}
\hfill
\begin{minipage}[t]{.45\linewidth}
\centering
\epsfig{file = 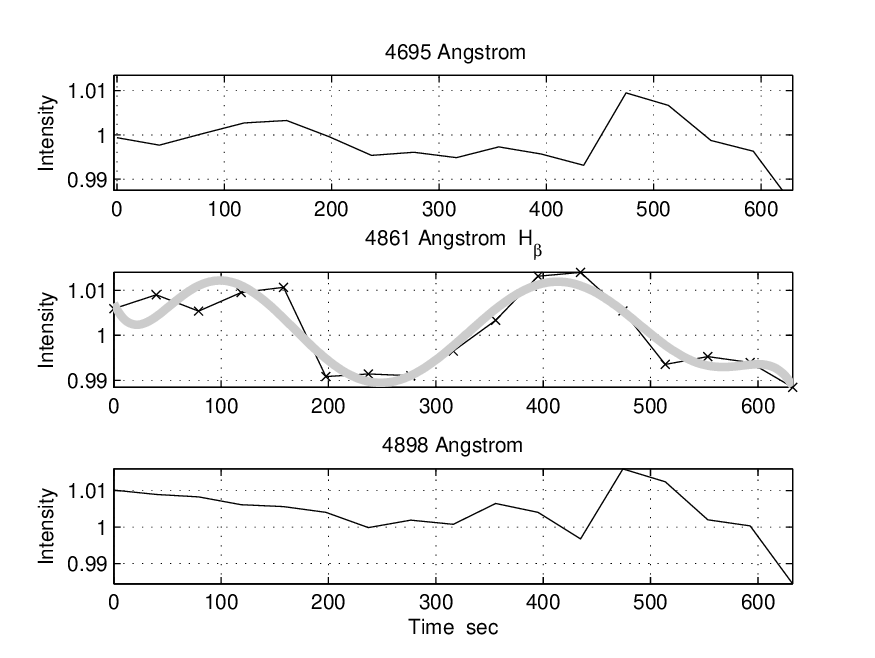,width = 1.0\linewidth} \caption{Light curves in the continuum (upper and lower panels) and in the $H_{\beta}$  line (middle panel).}\label{fig6}
\end{minipage}
\hfill
\end{figure}
%

\begin{figure}[!h]
\centering
\begin{minipage}[t]{.45\linewidth}
\centering
\epsfig{file = 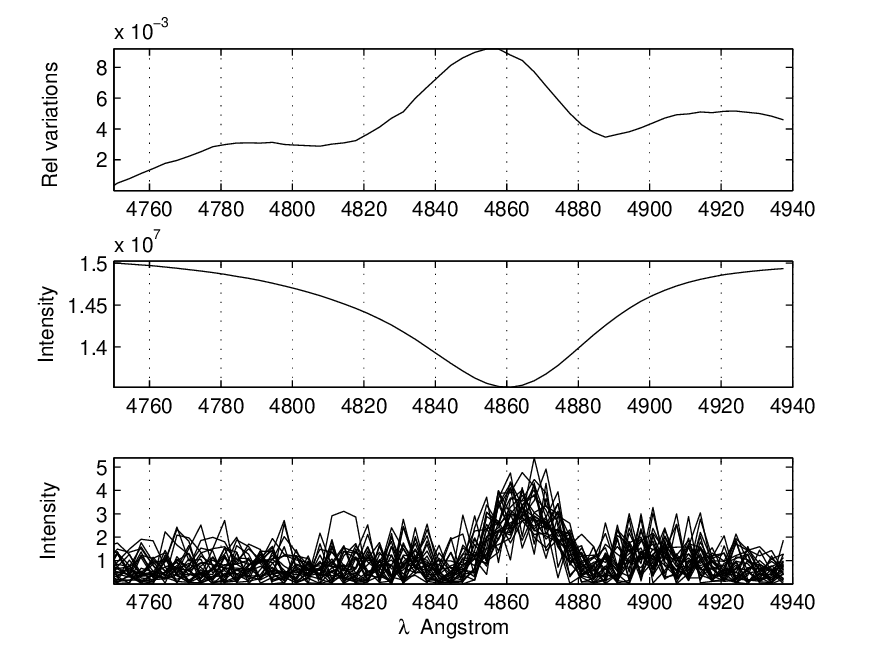,width = 1.0 \linewidth} \caption{Relative variability as a function of position in the $H_{\beta}$ line profile (top panel) and Zeeman signature of the line profile (low panel).}\label{fig7}
\end{minipage}
\hfill
\begin{minipage}[t]{.45\linewidth}
\centering
\epsfig{file = 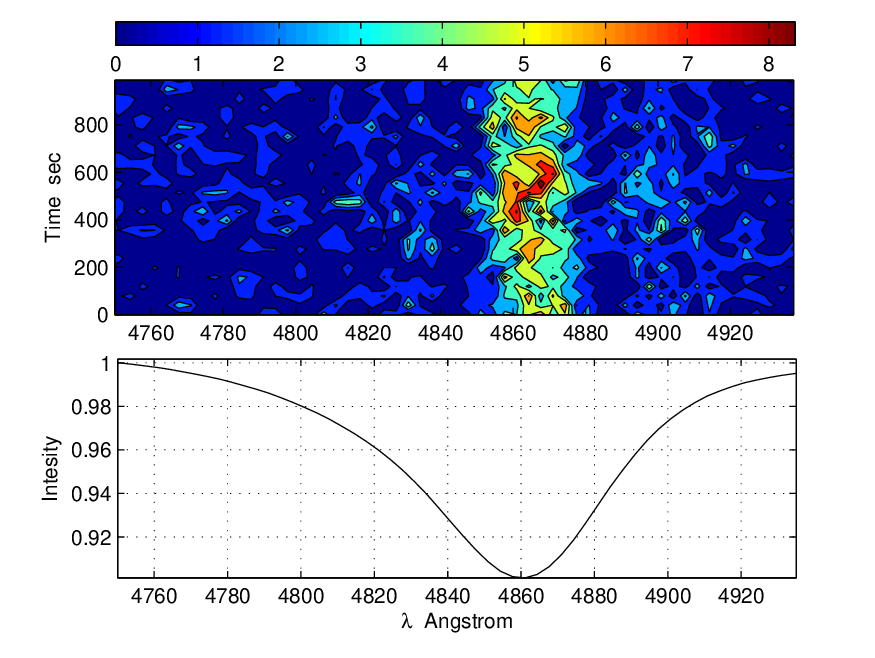,width = 1.0\linewidth} \caption{Oscillations restored in the  $H_{\beta}$ line.}\label{fig6}
\end{minipage}
\hfill
\end{figure}
\begin{figure}[!h]
\centering
\begin{minipage}[t]{.45\linewidth}
\centering
\epsfig{file = 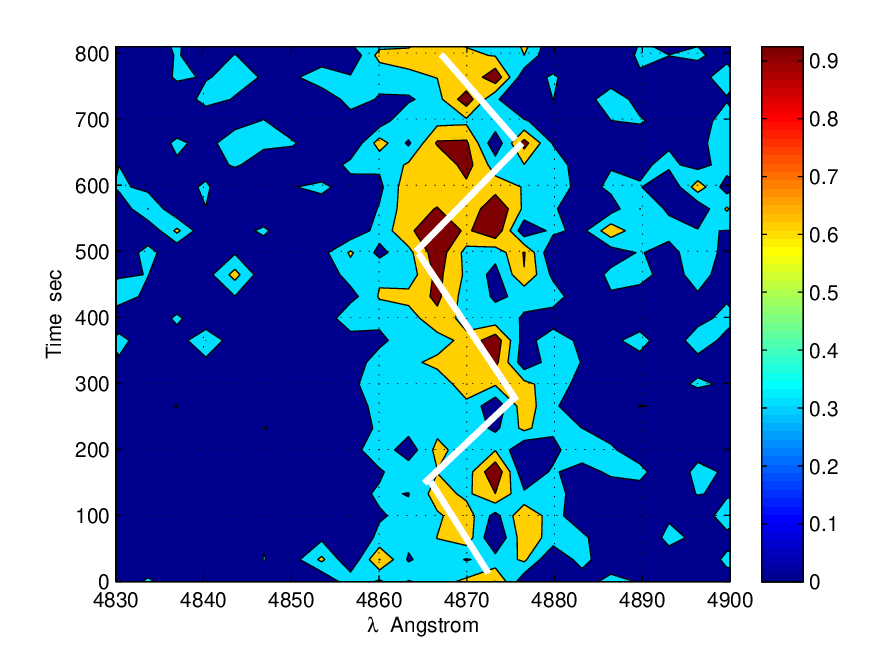,width = 1.0\linewidth} \caption{Variability in the $H_{\beta}$ line profile.}\label{fig8}
\end{minipage}
\hfill
\begin{minipage}[t]{.45\linewidth}
\centering
\epsfig{file = 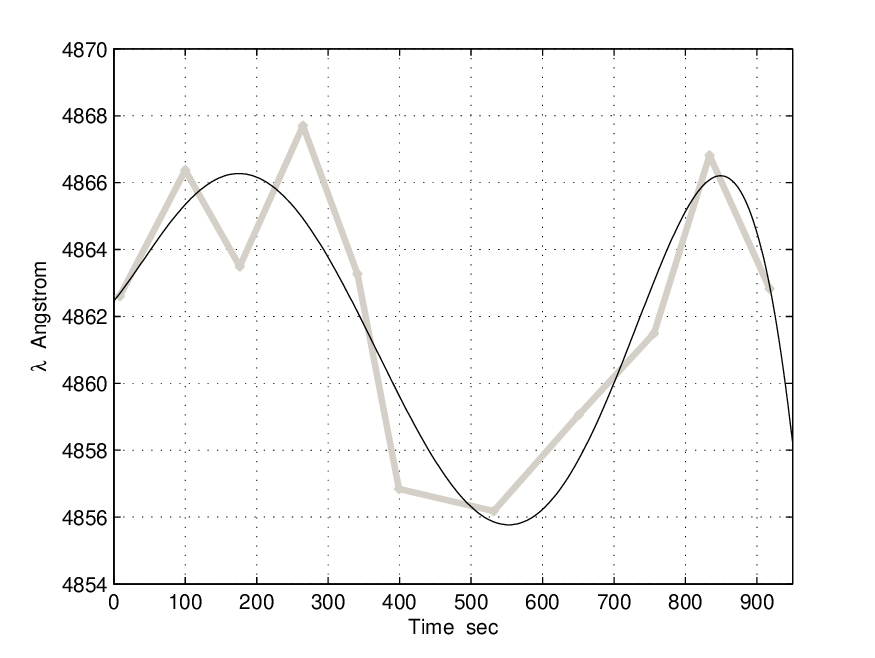,width = 1.0\linewidth} \caption{Variations in the profile of the $H_{\beta}$ line.}\label{fig8}
\end{minipage}
\hfill
\end{figure}
%


\section*{\sc Restoration of spectral line activity}
\indent \indent Qualitative analysis of spectrometric data consists of identifying peaks and estimating their position and intensity. Deconvolution methods are an effective tool for peak detection and estimation.

To estimate the dynamical spectrum of the nondadial oscillations, we use the technique of polarization estimation through an integral function called Zeeman signature Donati \cite{Donati}.
The new spectrum estimate obeys the expression

$$ I_{loc} =  |\frac{\partial^{2}I}{\partial^{2}\lambda} |  $$
where $I$ is the profile of the spectral line, $\lambda$ is the vawe length.
If the line profile is approximated by a Gaussian curve, this transformation preserves the line profile and increases the signal-to-noise.

Figure 11 shows the relative variability as a function of position in the $H_{\beta}$ line profile (upper panel) and the Zeeman signature of the line profile (lower panel).

The formula below deconvolves image $I$ using the maximum likelihood algorithm and an initial estimate of the point-spread function (PSF).

$$ I_{\lambda} = \int PSF\cdot I_{loc}(\lambda-\lambda_{c})\cdot \partial \lambda_{c}   $$

We use the Richardson-Lucy algorithm. It uses a statistical model to generate the data and is based on Bayes formula.
This iterative method forces the deconvolved spectra to be non-negative. The Richardson-Lucy iteration converges to a maximum likelihood solution for Poisson statistics in the data.

The image in Fig. 12 shows the restoration of the oscillation spectrum in the $H_{\beta}$ line after 200 iterations. The 'deconvblind' function in MATLAB returns the restored image $I$. We used a Gaussian model for the PSF peaks.

Figure 13 shows the temporal changes in the line profile after recovery. In Figure 14, the period of variations is about 600 seconds, and the amplitude is about 5 Angstroms. These quantities characterize the variations of the line profile in the observer's coordinate system. The variations of the radial velocity are shown in Fig. 6.

\begin{figure}[!h]
\centering
\begin{minipage}[t]{.45\linewidth}
\centering
\epsfig{file = 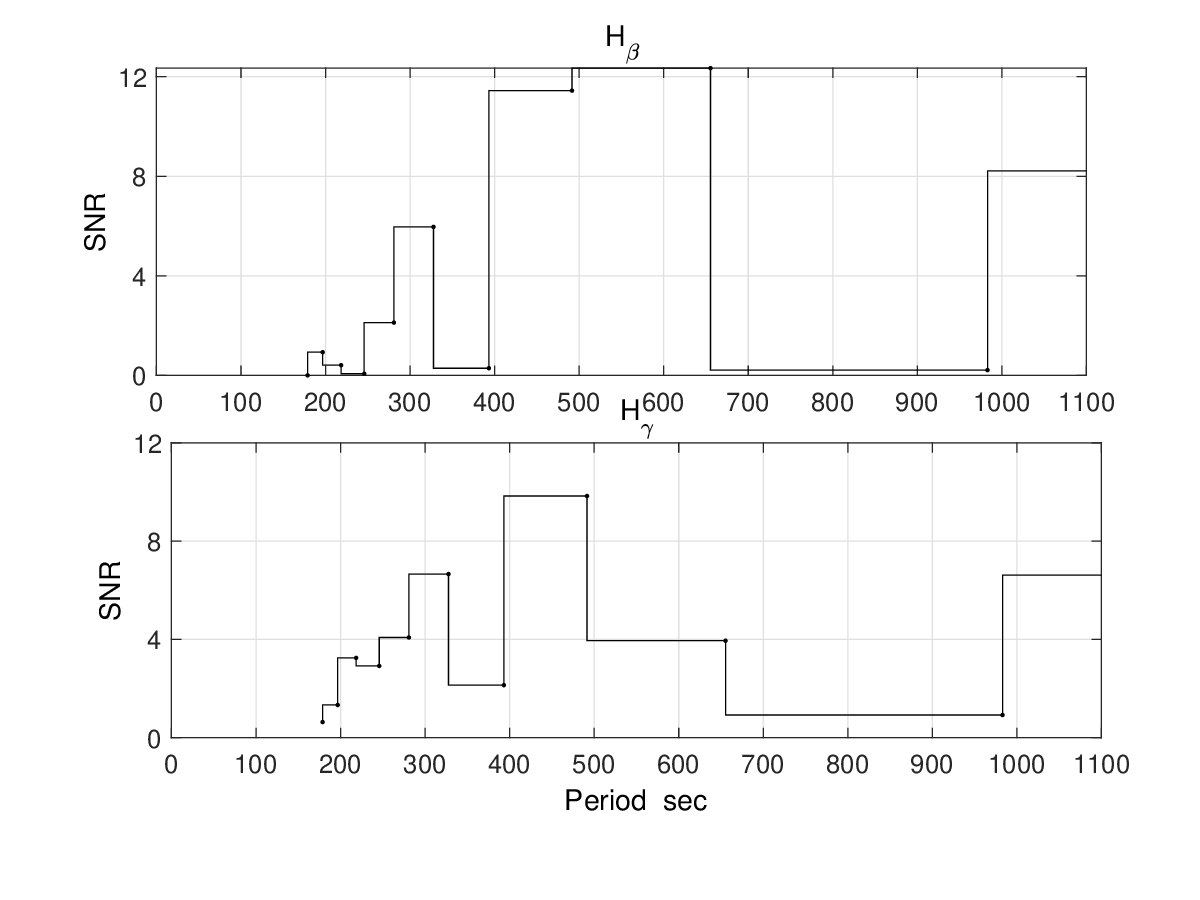,width = 1.0\linewidth} \caption{Fourier power spectrum. The first set. }\label{fig7}
\end{minipage}
\hfill
\begin{minipage}[t]{.45\linewidth}
\centering
\epsfig{file = 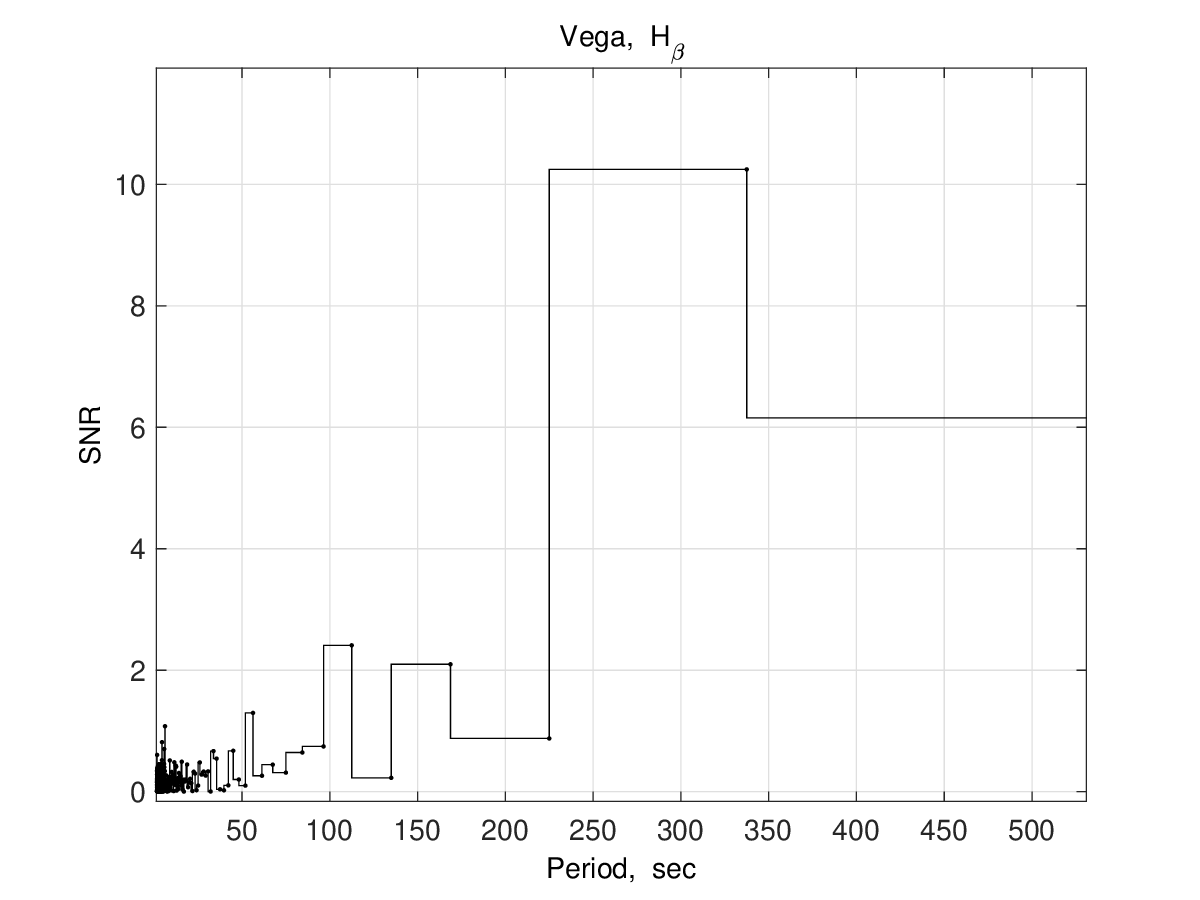,width = 1.0\linewidth} \caption{Fourier power spectrum. The second set.}\label{fig8}
\end{minipage}
\hfill
\end{figure}
%


\section*{\sc results and conclusions}

\indent \indent The power spectra of Vega in different observational sets show the same structure: (1) The same peak with a period of 300 seconds (Figs. 15 and 16). (2) The first set of observations with a longer duration shows multiple periods of 600 and 1200 seconds. (3) The power spectra in the H$_{\beta}$ and H$_{\gamma}$ lines are identical.

Fig. 15 demonstrates multiple periods of 600 and 1200 sec. Warner and Robinson \cite{Warner} assumed the equal spacing in frequency of periods to be due to the rotational splitting of nonradial modes.

The theory predicts that the visible star layers are driven by a set of standing waves, each covering the whole surface, and having a radial velocity $V$ whose temporal and spatial dependence
is given by 
$$V \sim \sin(\sigma \cdot t) \cdot Y^{l}_{m}(\theta,\phi)  $$
where $ Y^{l}_{m}(\theta,\phi)$ denotes the spherical harmonics, $\theta$ the colatitude, and $\phi$ the azimuth angle in the spherical coordinate, $\sigma $ the frequency, and $t$ the time \cite{Unno1979}.


The nonradial mode of oscillations of the spherical harmonics with quantum numbers $(k, l, m)$ and $m (= -l,..., 0, ..., l)$ are integers, for stars with angular rotation velocity $\Omega$ has the following form
$$ \sigma = \sigma_{0} +2\cdot m \cdot \Omega \cdot C_{k,l} $$
where $\sigma_{0}$ is the mode frequency in the absence of rotation, $C_{k,l}$ is a constant depending on the structure of the star. The constant $C_{k,l}$ as suggestid by Ledoux \cite{Ledoux} are small and $\ll$ 1. For the homogeneous model it is equal to unity. Then the frequency splitting has the form  \cite{Deubner} :
$$ \Delta\, \sigma = 2\cdot |m| \cdot \Omega$$

For a rotational frequency of 1/0.71 day$^{-1}$ and $\Delta\,\sigma $ = $2*\pi$/300 Hz, we obtain an estimate of the quantum number $m \simeq$ 15.

Acoustic p-modes occupy the high-frequency region with periods from 300 to 1200 s.
Horizontal wavelengths can be provided, respectively, by eigenmodes $l$ and $m \simeq$ 15, in which large amplitude spots appear to have average sizes of about 800 Mm for Vega. Note that the horizontal scale for oscillating features for the Sun is between 10 and 30 Mm \cite{Musman}.

Radial velocity $V$ according to Fig. 6 is about 0.3 Angstrom, which corresponds to 36 km/s.


The present work solves the problem of searching for the intrinsic variability of Vega's spectrum lines.

Dynamic spectroscopy of Vega using a spectrograph with a spectral resolution of R $\sim$ 600 allowed us to detect rapid variations of the hydrogen spectral lines. The characteristic time of the observed variations ranges from 300 to 1200 sec. The horizontal scale for oscillating elements is about 800 Mm, which is comparable to the solar radius. The radial velocity of the variations is about 36 km/s.



\end{document}